\documentclass[runningheads]{llncs}
\usepackage{graphicx}
\usepackage{blindtext}
\usepackage{hyperref}
\begin{document}
\title{SQA-SAM: Segmentation Quality Assessment for Medical Images Utilizing the Segment Anything Model\thanks{Work in progress. Email: yizhe.zhang.cs@gmail.com}}
%
%
\author{Yizhe Zhang\inst{1}\and
Shuo Wang\inst{2}\and Tao Zhou\inst{1}\and Qi Dou\inst{3}\and Danny Z. Chen\inst{4}}
\authorrunning{Y. Zhang et al.}
%
\institute{Nanjing University of Science and Technology, Nanjing, China\and
Digital Medical Research Center, School of Basic Medical Sciences, Fudan University, Shanghai, China
\and
The Chinese University of Hong Kong, Hong Kong, China
\and
University of Notre Dame, Notre Dame, USA}
\titlerunning{SAM for Medical Image Segmentation Quality Assessment}
\maketitle              
\begin{abstract}

Segmentation quality assessment (SQA) plays a critical role in the deployment of a medical image based AI system. Users need to be informed/alerted whenever an AI system generates unreliable/incorrect predictions. With the introduction of the Segment Anything Model (SAM), a general foundation segmentation model, new research opportunities emerged in how one can utilize SAM for medical image segmentation. In this paper, we propose a novel SQA method, called SQA-SAM, which exploits SAM to enhance the accuracy of quality assessment for medical image segmentation. When a medical image segmentation model (MedSeg) produces predictions for a test image, we generate visual prompts based on the predictions, and SAM is utilized to generate segmentation maps corresponding to the visual prompts. How well MedSeg's segmentation aligns with SAM's segmentation indicates how well MedSeg's segmentation aligns with the general perception of objectness and image region partition. We develop a score measure for such alignment. In experiments, we find that the generated scores exhibit moderate to strong positive correlation (in Pearson correlation and Spearman correlation) with Dice coefficient scores reflecting the true segmentation quality.
The code and updates will be made available at
\url{github.com/yizhezhang2000/SQA-SAM}.

\keywords{Medical Image Segmentation  \and Segmentation Quality Assessment \and Segment Anything Model \and Segmentation Quality Control.}
\end{abstract}
\section{Introduction}
Numerous deep learning (DL) based medical image segmentation models were developed in the last decade, driving state-of-the-art segmentation performance to higher and higher accuracy on test data. It is well-known that when test samples are i.i.d.~(independent and identically distributed) drawn from the sample distribution of the training data, modern DL-based segmentation models often perform quite well. But, in real world scenarios, the current best DL models still suffer from unreliable outcomes when encountering unfamiliar samples (those that are not i.i.d.~drawn from the sample distribution of the training data). More importantly, there is generally a lack of notion of segmentation quality when a segmentation model provides its predictions, leading to situations where medical practitioners lack confidence in using such medical AI systems in practice. 
 
Recently, more research emphasis has been put on developing trustworthy medical AI systems. Mehrtash et al.~\cite{mehrtash2020confidence} investigated the issue of poorly calibrated DL models for medical image segmentation, comparing cross-entropy and Dice loss functions and proposing a model ensemble approach for improved confidence calibration. Zhang et al.~\cite{zhang2022usable} proposed a novel evaluation metric that simultaneously examines both confidence calibration and segmentation accuracy to measure the usability and reliability of a medical image segmentation model. Zou et al.~\cite{zou2023review} contributed a comprehensive review of newly developed segmentation networks with uncertainty estimation mechanisms, exploring diverse methods such as ensemble learning, Bayesian inference, and Monte Carlo dropout. These techniques quantify model uncertainty, enabling informed decision-making and bolstering the overall trustworthiness of medical AI systems.

In this paper, we develop a novel SAM-based quality control approach, called SQA-SAM, for medical image segmentation. Instead of focusing on equipping segmentation networks with new internal uncertainty estimation mechanisms, we propose an external system to evaluate segmentation quality based on test images and the model's predictions. This system leverages the Segment Anything Model (SAM)~\cite{Kirillov-SAM-2023}, a general-purpose foundation segmentation model trained using 11 million images and over 1 billion masks. SAM excels at segmenting objects given point or bounding-box prompts. The medical image segmentation model under investigation (MedSeg) provides point and bounding-box prompts, which can then be used to generate new segmentation masks using SAM. The agreement between MedSeg's segmentation and SAM's segmentation indicates how well MedSeg's segmentation aligns with general perception. This alignment can then be used to derive a score for measuring the quality of MedSeg's segmentation output. 
Our experiments show that the scores thus computed exhibit moderate to strong positive correlation (in Pearson correlation and Spearman correlation) with Dice coefficient scores reflecting the true segmentation quality.

\begin{figure}[t]
\centering
\includegraphics[width=1.0\textwidth]{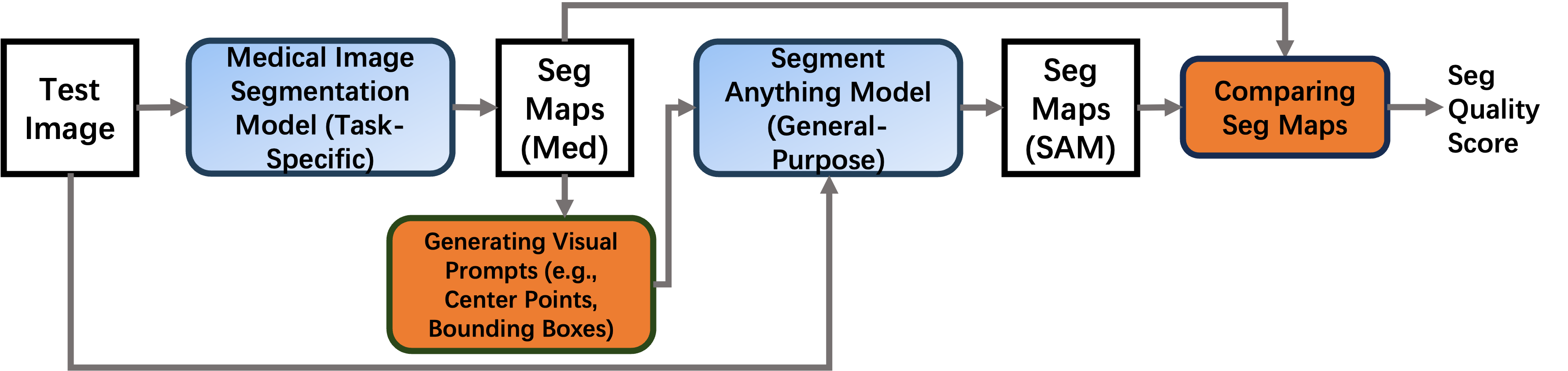}
\caption{Utilizing SAM for assessing segmentation quality for a medical image segmentation task.} \label{fig:result1}
\end{figure}

\section{Related Work}
We first provide a brief review of related work on segmentation quality assessment, categorized into two groups: methods that utilize labeled samples for training an SQA model and unsupervised methods that require no labeled samples to perform. Next, we present a review of recent methods that employ the Segment Anything Model (SAM) for improving medical image segmentation.

\subsection{Training an SQA Model Using Labeled Samples}
Huang et al.~\cite{huang2016qualitynet} proposed a learning-based approach for estimating segmentation quality using DL networks; three options were developed for constructing the network, mostly on where the segmentation mask is fused with the features/images in the process of estimating the segmentation quality score. In a similar fashion, Zhou et al.~\cite{zhou2020robust} proposed to use two sequential networks for SQA. Devries et al.~\cite{devries2018leveraging} utilized uncertainty maps to aid the estimation of the quality score; raw images, segmentation maps, and uncertainty maps are combined and fed to a DL-based network for generating a segmentation quality score. Rottmann et al.~\cite{rottmann2020prediction} proposed aggregating the dispersion of softmax probabilities to infer the true segmentation IoU. Rahman et al.~\cite{rahman2022fsnet} proposed using an encoder-decoder architecture to detect segmentation failure cases at the pixel level; multi-scale features from the segmentation network are extracted and fed to the decoder for generating a map that highlights the mis-classified/mis-segmented pixels. 

A common problem with learning-based SQA methods arises when test samples are not drawn from exactly the same distribution as the samples used when building the learning-based SQA model. The neural networks for estimating SQA might be familiar with certain types of errors that the segmentation model tends to produce, but when the test samples become unfamiliar, the accuracy of the learning-based SQA methods would drop significantly for assessing the segmentation quality.

\subsection{SQA That Requires No Additional Labeled Samples}
Zhang et al.~\cite{zhang2008image} presented an extensive survey of unsupervised segmentation quality evaluation methods. The early unsupervised SQA approaches primarily centered around the formulation of manually crafted features, including but not limited to gray-scale differences, intra-region uniformity, busyness, and shapes. These features were employed to gauge segmentation quality in the absence of ground truth labels. Recently, Chen et al.~\cite{chen2023evaluation} utilized 14 pre-defined metrics under different assumptions to evaluate the 
quality of cell segmentation. After attaining the evaluation results of the pre-defined metrics, they further performed PCA (Principal Component Analysis) to obtain a single score describing the segmentation quality. Audelan et al.~\cite{audelan2021unsupervised} estimated segmentation quality by comparing each segmentation with the output of a probabilistic segmentation model that relies on intensity and smoothness assumptions. While many handcrafted metrics successfully captured various facets of segmentation quality, it became evident that, in more realistic and complex scenarios, handcrafted simple measures are inadequate for comprehensive assessment of segmentation quality.

\subsection{Utilization of SAM for Medical Image Segmentation}
Since the introduction of SAM~\cite{Kirillov-SAM-2023}, many attempts have been made in utilizing SAM for medical image segmentation. Ma et al.~\cite{ma2023segment} studied fine-tuning SAM for medical images using a large collection of medical images with segmentation annotations. Huang et al.~\cite{huang2023segment} tested the performance of SAM for medical image segmentation using 53 open-source datasets and showed some limitations of SAM when directly applying it to medical images. Zhou et al.~\cite{zhou2023can} tested the performance of SAM for polyp segmentation and compared it with state-of-the-art polyp segmentation models. Zhang et al.~\cite{zhang2023input} proposed to use SAM for augmenting input images for medical image segmentation. SAM was also used for semi-supervised learning of medical image segmentation~\cite{zhang2023samdsk}. Previous studies demonstrated that SAM can provide a general perception of objectness, making it a valuable tool for improving medical image segmentation performance.

\section{Method}
\subsection{Objective}

The ultimate goal of segmentation quality assessment is to generate a score for a given test sample which is equivalent to the true quality score obtained by comparing the segmentation with the ground truth. For example, suppose the Dice coefficient is used for measuring the segmentation quality; the goal would be to generate/predict the Dice coefficient score given segmentation results and their corresponding raw images. Since this goal is very challenging, in this paper, we tackle a slightly easier task: Given a set of test samples with the predicted segmentation maps, generate SQA scores with high linear correlation and rank correlation with the true evaluation metric scores.

\subsection{Utilizing SAM for SQA}
Given a test image $x$, we apply a model $f$ to $x$ to obtain a segmentation map $\hat{y} \in \{0,1\}^{w\times h\times C}$, where $C$ is the number of classes of the segmentation task, and $w$ and $h$ denote the width and height of the input image. We then extract objects using a connected component (CC) algorithm from $\hat{y}$. For each class $i$, $i=1,2, \dots, C$, we apply the CC algorithm to the $i$-th channel of $\hat{y}$ and extract a set of objects corresponding to class $i$, denoted as $\mathcal{M}_i=\{m_1^i, m_2^i, \dots, m_{p_i}^i\}$, where each $m^i_j$
is a binary mask corresponding to one object, and $p_i$ is the number of objects in the $i$-th class map. We repeat this process for all the classes and obtain $\mathcal{M}_1, \mathcal{M}_2, \dots, \mathcal{M}_C$.

For each object $m_{j}^i \in \mathcal{M}_i$, we obtain all the pixel locations with pixel values equal to 1, and compute their center point and bounding box. We separately feed the center point and bounding box as prompts (together with the image $x$) to SAM, and obtain two binary masks: one induced by the center point prompt (denoted as $pm_{j}^i$) and the other induced by the bounding-box prompt (denoted as $bm_{j}^i$). We then calculate a score for object $m_{j}^i$ as $s_{j}^{i} = \tau (m_{j}^i, pm_{j}^i) + \tau (m_{j}^i, bm_{j}^i)$. Here, $\tau$ is a segmentation evaluation metric, and by default, we set it as the Dice coefficient metric. We repeat the above process for all the objects in $\mathcal{M}_i$, for $i=1, 2, \dots, C$. Finally, we compute an overall segmentation quality score ${s}$ for the segmentation map $\hat{y}$ as: 
\begin{equation}
s=\frac{\sum_{i=1}^C\sum_{j=1}^{p_i} s_{j}^{i}}{\sum_{i=1}^C\sum_{j=1}^{p_i}1}.
\label{eq-SQA-score}
\end{equation}

\subsection{Limitations}

We should note that the above computed scores do not address errors caused by incorrect class-label assignment. For example, if the segmentation masks are of good quality (e.g., with correct shapes) but the corresponding classes are assigned incorrectly, then the computed score in the above scheme is not able to capture such types of errors, as SAM is class-agnostic in general. Furthermore, if an object of interest is missed in the segmentation output of $f$, then the computed score would not be able to capture such missed detection cases, since the visual prompts are generated depending on the segmentation output to provide the initial pixel regions.

\section{Experiments}
We empirically study whether the generated segmentation quality scores have a positive correlation with the Dice coefficient scores (computed by comparing segmentation maps with ground truths). We use two datasets for the experiments, Polyp Segmentation in Endoscopic Images~\cite{fan2020pranet} and Retinal OCT Fluid Segmentation (RETOUCH)~\cite{bogunovic2019retouch}. Both these two datasets consider only one type of objects (e.g., polyp) and are binary segmentation tasks. We setup a baseline (Model Confidence) which is built upon the pixel-level confidence of the segmentation prediction: We take the max value of the prediction logits (after softmax) for each pixel, and
then takes the average confidence across all the pixels for each test sample.

\subsection{Polyp Segmentation in Endoscopic Images}
On the Polyp Segmentation in Endoscopic Images dataset~\cite{fan2020pranet}, we utilize the well-trained HSNet by Zhang et al.~\cite{zhang2022hsnet} for the polyp segmentation experiments. The test set comprises samples from five datasets. For simplicity and clarity, we report the performance of the segmentation quality assessment for all the test samples in one test run.

Table~\ref{table:sqa_polyp} reports the SQA scores computed by our proposed SQA-SAM method, showing a moderate to strong correlation with the Dice coefficient scores computed by comparing the segmentation results against ground truths. Furthermore, we assess the accuracy of using the SQA scores to detect samples with Dice coefficient scores in the bottom $k\%$ among all the test samples. In Table~\ref{table:accu_polyp}, we show the accuracy of the SQA scores in detecting samples of poor segmentation quality (with low Dice coefficient scores) when setting $k\%$ as 25\% and 50\%. Additionally, we investigate whether SAM can enhance segmentation output by replacing the output of HSNet with SAM's output. We find that 266 out of 798 samples receive improvement in segmentation quality (in Dice coefficient) after using SAM's segmentation maps instead of HSNet's output, but the other 532 samples actually exhibit degradation in segmentation quality. This demonstrates that simply utilizing SAM's output to replace the original segmentation model's output may not consistently result in improved segmentation output, echoing previous findings in~\cite{zhou2023can}.

\begin{table}[t]
\centering
\caption{Correlation between SQA scores and true Dice coefficient scores for polyp segmentation (higher numbers indicate better quality assessment performances).}
\begin{tabular}{|c|c|c|}
\hline
SQA Method & Pearson Corr. & Spearman Corr. \\\hline
{Model Confidence} & 0.414 & 0.096 \\
\hline
{SQA-SAM (ours)} & 0.518 & 0.659 \\
\hline
\end{tabular}
\label{table:sqa_polyp}
\end{table}

\begin{table}[t]
\centering
\caption{Accuracy of detecting samples with low Dice coefficient scores on the Polyp Segmentation in Endoscopic Images dataset (higher numbers indicate better performances). }
\begin{tabular}{|c|c|c|}
\hline
SQA Method  & Bottom 25\%  & Bottom 50\%  \\\hline
{Model Confidence} & 49.1\% & 50.8\%\\
\hline
{SQA-SAM (ours)} & 68.2\% & 75.9\%\\
\hline
\end{tabular}
\label{table:accu_polyp}
\end{table}

\subsection{Retinal OCT Fluid Segmentation}
RETOUCH~\cite{bogunovic2019retouch} is a benchmark and challenge dataset for designing and evaluating algorithms for detecting and segmenting various types of fluids in retinal optical coherence tomography (OCT) images. It provides representative OCT data across different retinal diseases, OCT vendors, and fluid types. We train the UNet++ model \cite{Zhou2018UNET-Plus} using the training set of this dataset and use the test set with images from the ``SPECTRALIS OCT'' vendor for testing the performances of quality assessment methods. Table~\ref{table:sqa_oct} shows that our QA-SAM method yields scores with a moderate to strong correlation with the true Dice coefficient scores. Table~\ref{table:accu_oct} shows the detection accuracy of using the segmentation quality scores to detect test samples with poor segmentation quality.

\begin{table}[t]
\centering
\caption{Correlation between SQA scores and true Dice coefficient scores for retinal OCT fluid segmentation (higher numbers indicate better quality assessment performances).}
\begin{tabular}{|c|c|c|}
\hline
SQA Method & Pearson Corr. & Spearman Corr. \\\hline
{Model Confidence} & 0.211 & 0.229 \\
\hline
{SQA-SAM (ours)} & 0.611 & 0.710 \\
\hline
\end{tabular}
\label{table:sqa_oct}
\end{table}

\begin{table}[t]
\centering
\caption{Accuracy of detecting samples with low Dice coefficient scores on the RETOUCH dataset (higher numbers indicate better performances). }
\begin{tabular}{|c|c|c|}
\hline
SQA Method  & Bottom 25\%  & Bottom 50\%  \\\hline
{Model Confidence} & 32.7\% & 59.3\%\\
\hline
{SQA-SAM (ours)} & 67.3\% & 74.8\%\\
\hline
\end{tabular}
\label{table:accu_oct}
\end{table}

\section{Conclusions}
In this paper, we proposed a new effective method, SQA-SAM, which utilizes the Segment Anything Model (SAM) for segmentation quality assessment on medical images. SAM provides a fruitful way to examine how well segmentation results align with the general perception, and such alignment can indicate the segmentation quality. Experiments on two medical image segmentation tasks demonstrate the effectiveness of our proposed method. Further research will be conducted on more medical image segmentation tasks and datasets, and on further refining and extending the proposed SQA method.

\bibliographystyle{plain}
\bibliography{ref}
\end{document}